\documentclass[letterpaper,prl,twocolumn,showpacs,superscriptaddress]
{revtex4}

\usepackage{graphicx}

\newcommand{\avg} {\langle g\rangle}

\newcommand{\avt} {\langle t\rangle}
\newcommand{\avT} {\langle T\rangle}
\newcommand{\cN} {{\cal N}}
\newcommand{\eg} {{e.g., }}

\newcommand{\ie} {{i.e., }}

\begin{document}

%========================================================================
\title
{Chain-reaction cascades in surfactant monolayer buckling}
%========================================================================

\author{Ajaykumar Gopal}
\affiliation{Department of Chemistry and James Franck Institute,
University of Chicago, Chicago, Illinois 60637}
%%\email{bcui@stanford.edu}

\author{Vladimir A.\ Belyi}
\affiliation{Department of Physics and James Franck Institute,
University of Chicago, Chicago, Illinois 60637}

\author{Haim Diamant}
\email{hdiamant@tau.ac.il}
\affiliation{School of Chemistry, Raymond and Beverly Sackler Faculty
of Exact Sciences, Tel Aviv University,
Tel Aviv 69978, Israel}

\author{Thomas A.\ Witten}
\affiliation{Department of Physics and James Franck Institute,
University of Chicago, Chicago, Illinois 60637}

\author{Ka Yee C.\ Lee}
\affiliation{Department of Chemistry and James Franck Institute,
University of Chicago, Chicago, Illinois 60637}

\date{September 6, 2004}

\begin{abstract}
  Certain surfactant monolayers at the water--air interface have
  been found to undergo, at a critical surface pressure, a
  dynamic instability involving multiple long folds of micron width.
  We exploit the sharp monolayer translations accompanying folding
  events to acquire, using a combination of fluorescence microscopy
  and digital image analysis, detailed statistics concerning the
  folding dynamics.  The motions have a broad distribution of
  magnitudes and narrow, non-Gaussian distributions of angles and
  durations. The statistics are consistent with the occurrence of
  cooperative cascades of folds, implying an autocatalytic process
  uncommon in the context of mechanical instability.
\end{abstract}


\maketitle
%------------------------------------------------

Surfactant monolayers are found in many systems containing water--air
or water--oil interfaces where surface tension, wetting, or
liquid-film stability are to be controlled \cite{Birdi}.  In recent
years a remarkable variety of three-dimensional structures have been
discovered upon lateral compression of surfactant monolayers,
including straight folds
\cite{KYLee,Ajay,Knobler02}, convoluted folds
\cite{Knobler02,Bruinsma}, and attached vesicular objects of various
shapes \cite{Ajay}.  These instabilities have distinctive length
scales ranging between 0.1 and 10 $\mu$m.  Thus the
predominant stress relaxation is neither at the molecular level, which
would lead to breakage or dissolution \cite{breakage}, nor at the
macroscopic one, which would lead to long-wavelength buckling
\cite{buckling}.  Moreover, the transitions occur under a net {\it
  tensile} stress, \ie at surface pressures smaller than the bare
surface tension of water (72 mN/m).  Though the actual relaxation
mechanism is unknown, a plausible driving force may be bilayer
cohesion, \ie the preference of the hydrophobic surfactant tails to
join rather than remain in contact with air \cite{Bruinsma}. Domain
boundaries in biphasic monolayers or grain boundaries in monophasic
ones entail nanoscale topographies, which should cause localized
buckling at low enough, positive tension \cite{mesa}, thus lowering
the nucleation barrier for bilayer cohesion.

\begin{figure}[tbh]
\centerline{\resizebox{0.45\textwidth}{!}
{\includegraphics{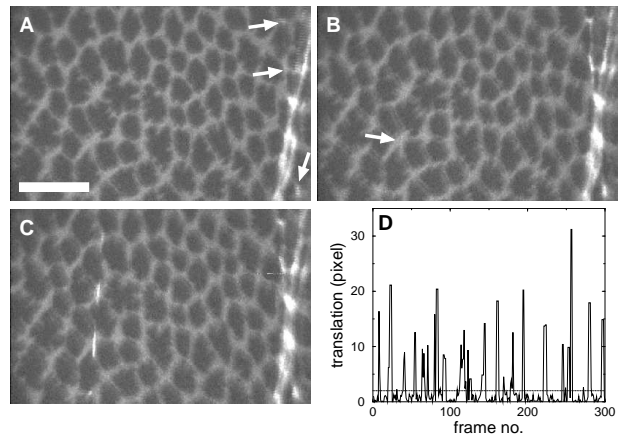}}}
\caption[]
{(a)--(c) Fluorescence micrographs separated by 1/30 s intervals,
  showing the nearly-simultaneous formation of two folds. The images
  are blurred by monolayer motion. The scale bar
  length is 50 $\mu$m. (d) Typical output of the tracking
  program, showing the monolayer translation within the field of view
  in sequential video frames.  The spikes correspond to folding events
  occurring out of view.  The dotted line shows the threshold used for
  event identification.}
\label{fig_fold}
\end{figure}

Out of the newly observed structures, the straight folds stand out as
essentially different, corresponding to a more solid-like,
nonequilibrium response of the monolayer
\cite{KYLee,mesa,Ajay,Knobler02}. They are observed at lower
temperature and higher compression rate, occur at higher critical
pressures ($\sim$ 70 mN/m), and are anisotropic, \ie aligned on
average perpendicular to the compression direction.  A fold comprises
a piece of monolayer of micron width and macroscopic length probably
bound into a bilayer strip (Fig.\ \ref{fig_fold}).  The main obstacle
in studying fold formation is that it is a nucleated event initiated
at an unpredictable spot and lasting a fraction of a second. Very
rarely is such an event captured inside the field of view as in Fig.\ 
\ref{fig_fold}.  Nevertheless, whenever a fold forms elsewhere, the
viewed piece of monolayer translates sharply and uniformly.  Watching
the monolayer jump as a result of such events, one is struck by the
uniformity and unidirectionality of the motion, in contrast with the
monolayer heterogeneity. The statistical analysis presented
below corroborates this impression, revealing such features as
anomalously narrow distributions of translation angle and
duration.

Insoluble monolayers containing a 7:3 mixture of
dipalmitoylphosphocholine (DPPC) and palmitoyloleoylphosphoglycerol
(POPG) were spread on the air--water interface from a chloroform
solution at $25^\circ$C. This was done on a rectangular Langmuir
trough of maximum area $145$ cm$^2$, fitted with two mobile Teflon
barriers of length $6.35$ cm each, allowing for symmetric lateral
compression.  The phospholipids formed a low-density
monolayer (area per molecule $a>100$ \AA$^2$), exhibiting coexistence
of two-dimensional gas and liquid phases.  Lateral compression led, at
$a\simeq 67$ \AA$^2$, to a homogeneous liquid phase followed, for
$a\lesssim 65$ \AA$^2$, by nucleation and growth of flower-like
condensed domains (Fig.\ \ref{fig_fold}) until, at $a\simeq 30$
\AA$^2$, folding began.  The relative barrier velocity was $0.1$ mm/s,
corresponding to compression rate of $6.35$ mm$^2$/s and strain rate
at the onset of folding of $0.00154$ s$^{-1}$.  Throughout the
compression the morphology was observed using epifluorescence video
microscopy and the surface tension monitored by a Wilhelmy surface
balance. Further details of the apparatus can be found in Ref.\ 
\cite{Ajay}.
The analog video was digitized into a series of 8-bit grayscale
bitmaps of $640\times 480$ pixels at a rate of $29.97$ frames/s.  The
series of images were then analyzed using a custom-made tracking program
whose algorithm will be detailed elsewhere \cite{ournext}.  A typical
output is shown in Fig.\ \ref{fig_fold}(d).  Using a velocity
threshold of 2 pixels per frame we identified 1817 events, recording
for each its starting time, duration and total $\vec{l}=(l_x,l_y)$
translation.

\begin{figure}[tbh]
\centerline{
\resizebox{0.31\textwidth}{!}
{\includegraphics{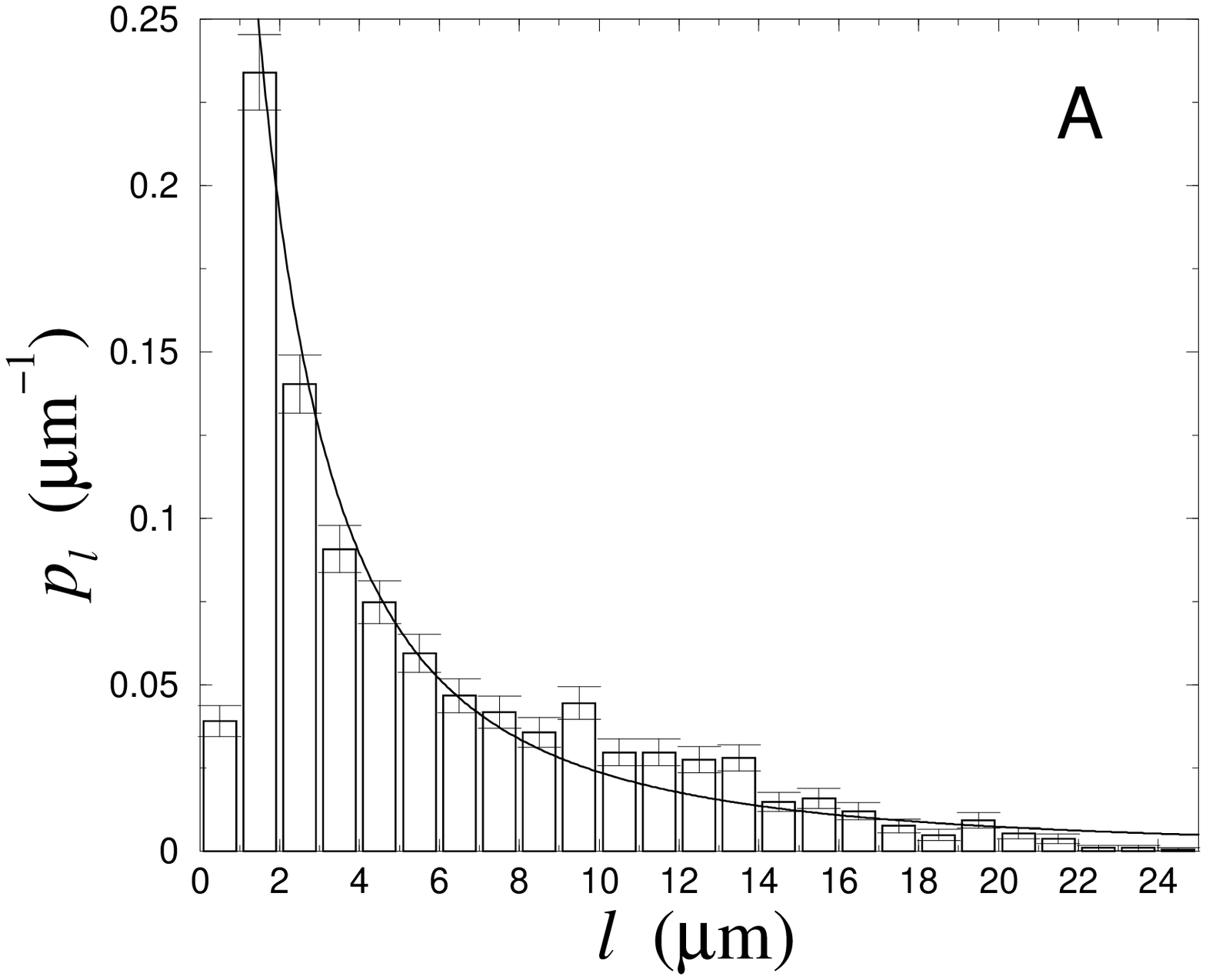}}}
\centerline{\resizebox{0.31\textwidth}{!}
{\includegraphics{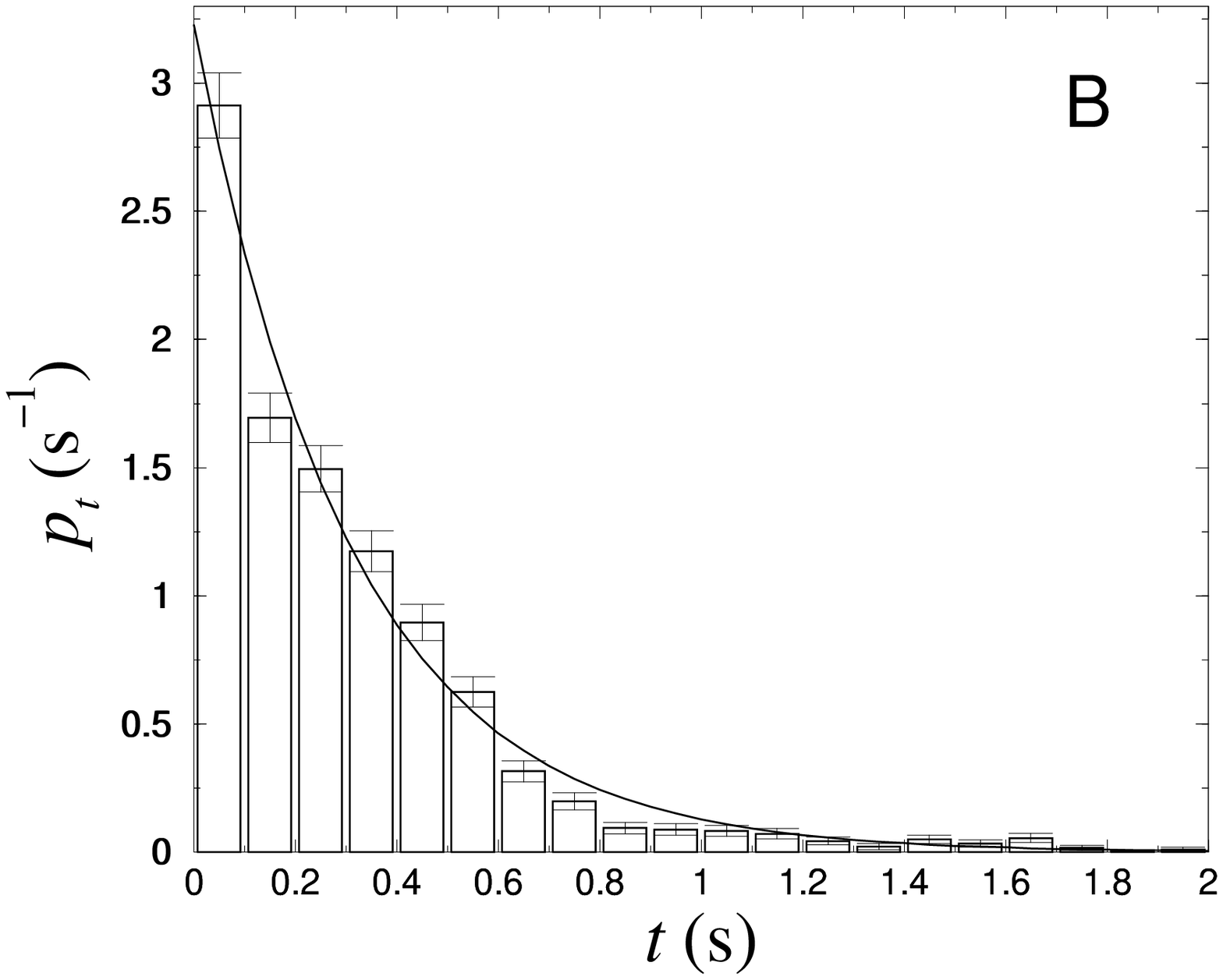}}}
\caption[]
{(a) Distribution of translations. The solid line is obtained assuming
  a cascade mechanism [Eq.\ (\ref{eq_pn})]. (b) Distribution of
  off-times. The solid line shows an exponential distribution using
  the measured mean off-time $\avt=0.31$ s.}
\label{fig_offtime_trans}
\end{figure}

Figure \ref{fig_offtime_trans}(a) shows the distribution of
translations, exhibiting a broad tail---translations ten times the
most probable one were observed. The distribution of off-times
$t$ (waiting times between events) is presented in Fig.\ 
\ref{fig_offtime_trans}(b).  The mean off-time is $\avt=0.31\pm 0.01$
s, \ie there are about three events per second.  The histogram fits
well an exponential distribution, $p_t(t) = \avt^{-1}e^{-t/\avt}$,
consistent with a Poissonian, uncorrelated sequence of events.  This
conclusion is strengthened by a lack of correlation between $l$ and
the off-times before or after the event.

\begin{figure}[tbh]
\centerline{\resizebox{0.32\textwidth}{!}
{\includegraphics{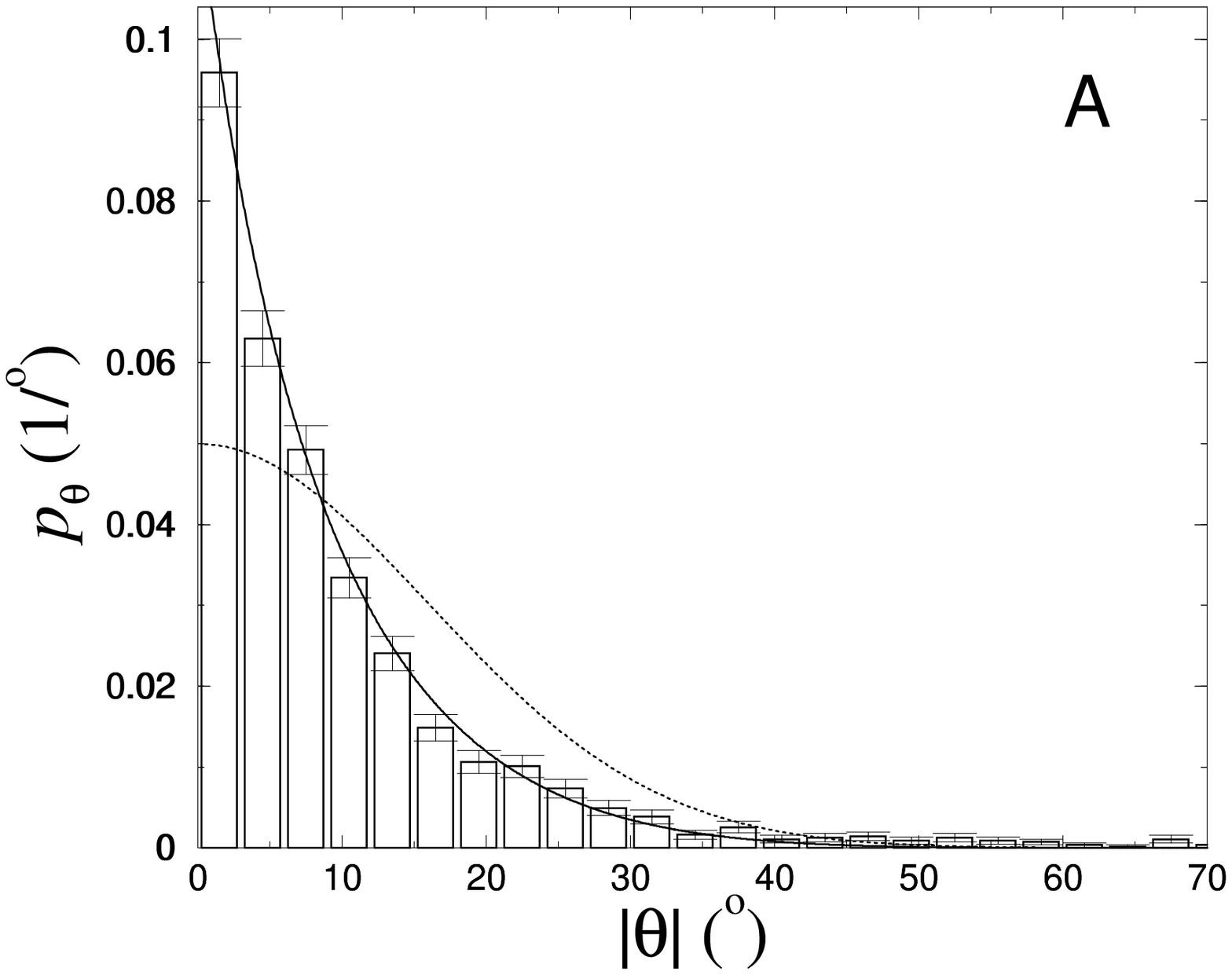}}}
\centerline{
\resizebox{0.31\textwidth}{!}
{\includegraphics{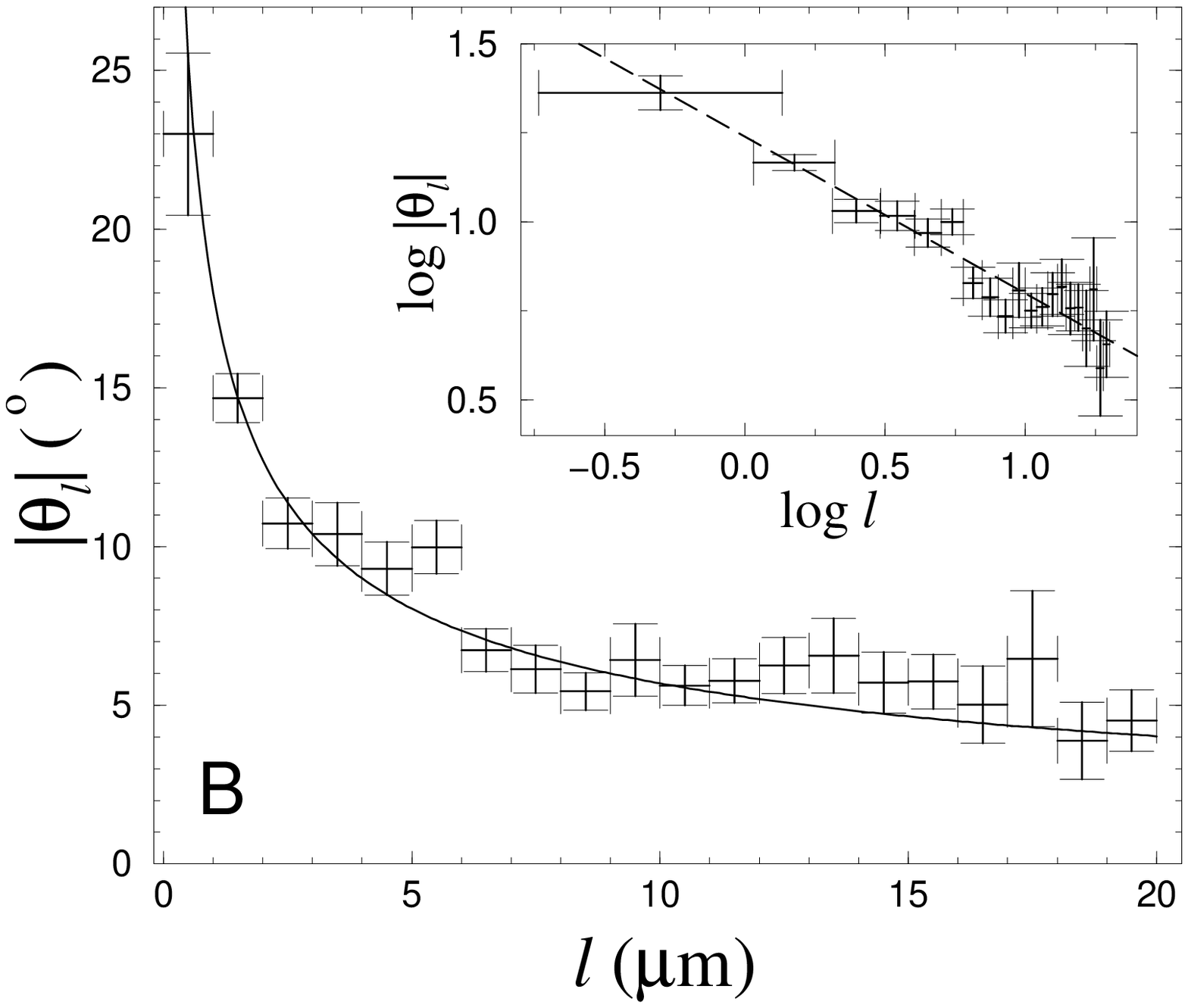}}}
\caption[]
{(a) Distribution of translation angles. The dotted line shows a
  Gaussian distribution with the measured $\langle\theta^2\rangle$.
  The solid line is obtained assuming a cascade mechanism [Eq.\ 
  (\ref{eq_p_theta})] with $\sigma_{\theta 1}=16^\circ$.  (b) Average
  angle of events whose translations fall within the same narrow
  range. The solid line is a fit to $|\theta_l|=\gamma l^{-1/2}$ [Eq.\ 
  (\ref{eq_theta_l})] with $\gamma=18\pm 3$ $(^\circ)\mu$m$^{1/2}$.
  The inset presents the same data on a log--log scale, the dashed
  line being an error-weighted linear fit with a slope of $-0.44\pm
  0.05$.}
\label{fig_angle}
\end{figure}

Figure \ref{fig_angle}(a) shows the distribution of translation angles
$|\theta|$, where $\tan\theta=l_y/l_x$, and $\theta=0$ corresponds to
motion parallel to the compression direction.  The distribution is
sharply peaked at $\theta=0$, with standard deviation
$\sigma_\theta\equiv\langle\theta^2\rangle^{1/2}=16.0\pm 0.3^\circ$.
Folding is thus highly anisotropic, implying an elastic response of
the monolayer within the folding time scale.  This is in line with the
viscoelasticity revealed by surface-rheology measurements in similar
systems \cite{Mohwald,Bruinsma}, which yields relaxation
times of order tens of seconds.
One would expect a Gaussian distribution of fold angles---because of
either the combined effect of many scattering factors or a Boltzmann
factor for the nucleation of a slanted fold with respect to the
direction of maximum stress, whose energy increases as
$\sin^2\theta\simeq\theta^2$.  The peak of the measured distribution,
however, is much sharper and clearly cannot be fitted by a Gaussian
distribution (dotted curve).  In Fig.\ \ref{fig_angle}(b) we show the
mean angle $|\theta_l|$ of events as a function of their translation
$l$. The two quantities are anticorrelated---larger events have
smaller angles.

\begin{figure}[tbh]
\centerline{\resizebox{0.31\textwidth}{!}
{\includegraphics{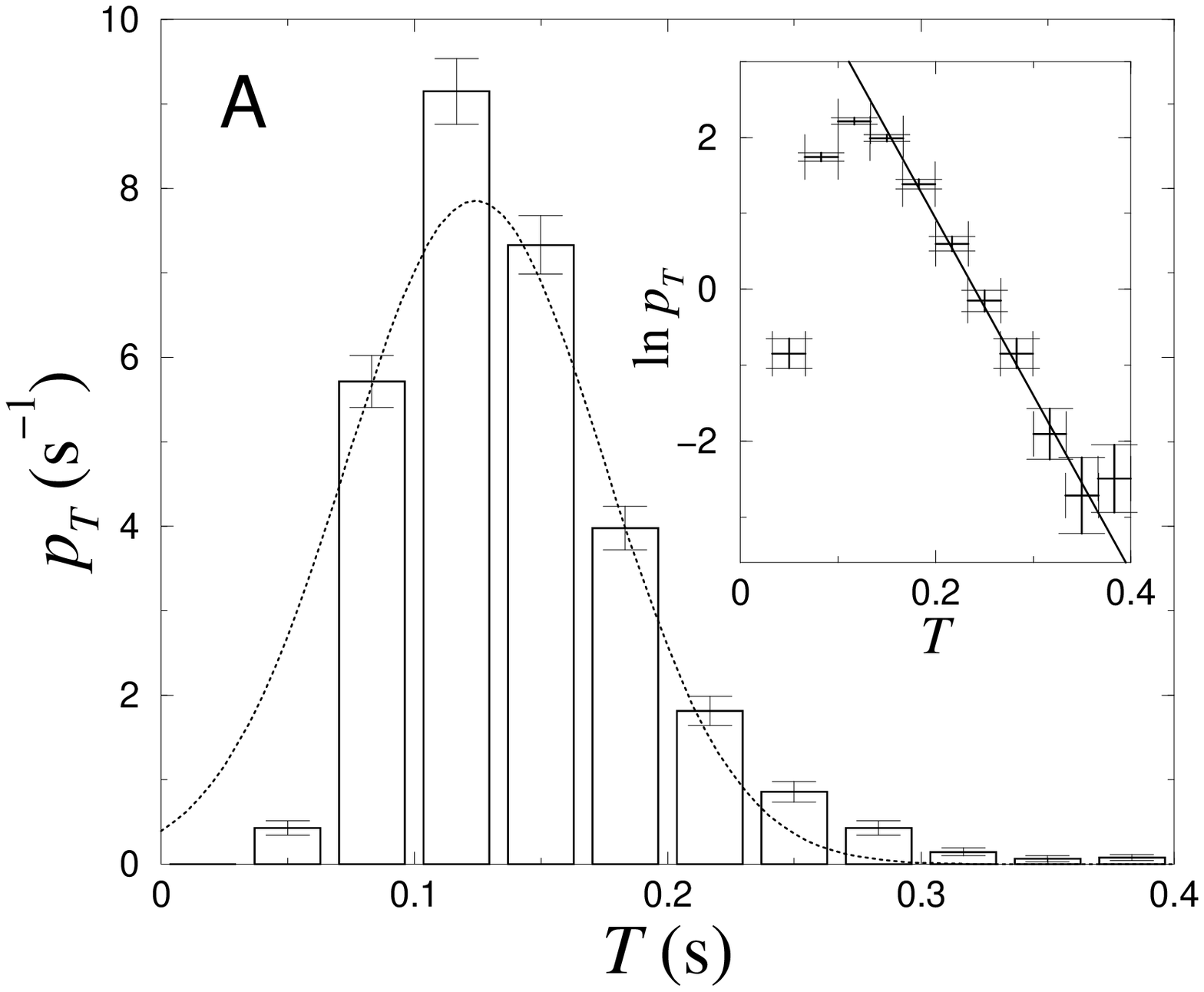}}}
\centerline{
\resizebox{0.32\textwidth}{!}
{\includegraphics{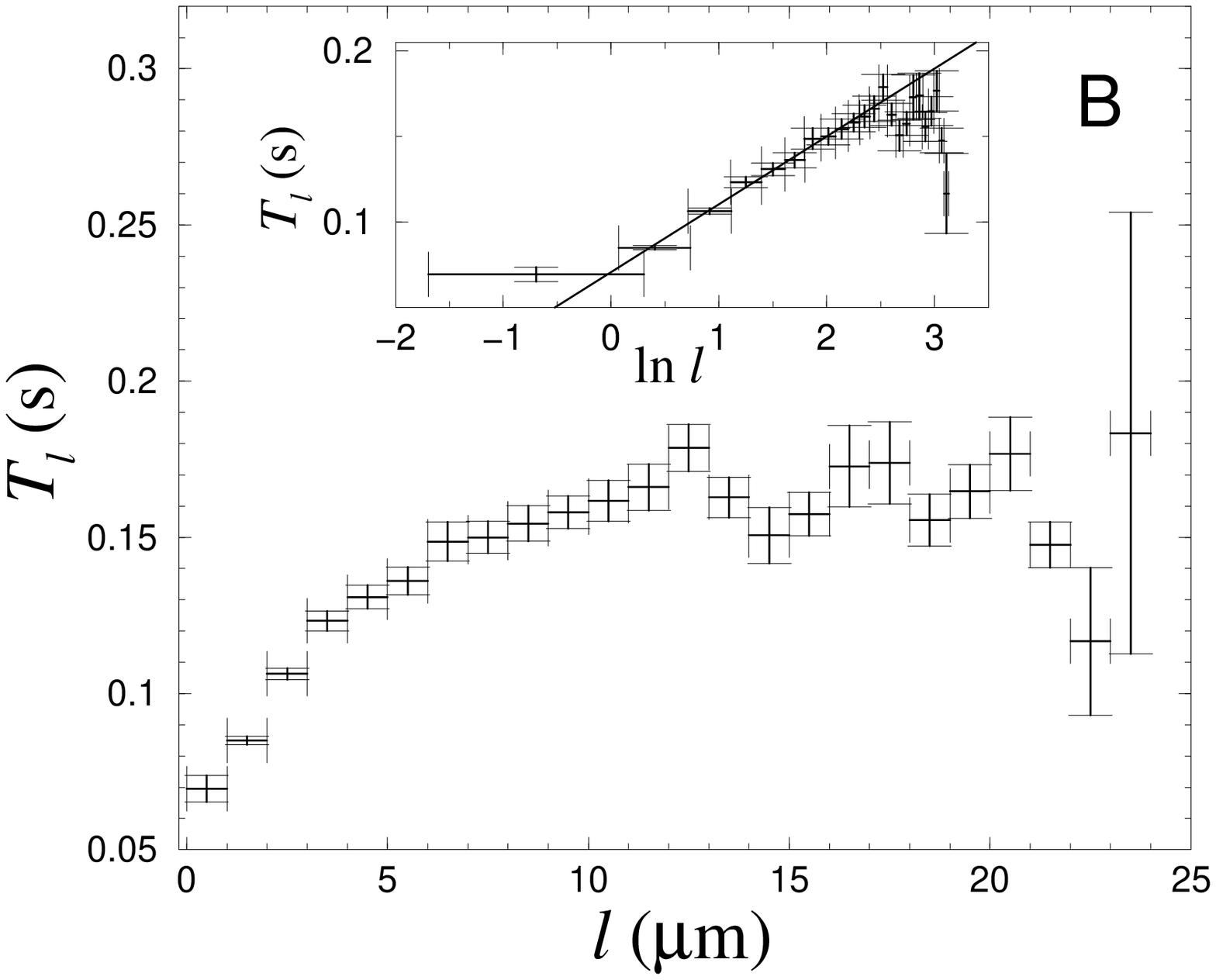}}}
\caption[]
{(a) Distribution of on-times.  The dotted line
  shows a Gaussian distribution using the measured mean and standard
  deviation. The inset presents the data on a linear--ln scale, the
  solid line being an error-weighted linear fit with a slope of
  $-23.3\pm 3$ s$^{-1}$.  (b) Average on-time of events whose
  translations fall within the same narrow range. The inset presents
  the same data on a ln--linear scale, the solid line being an
  error-weighted linear fit with a slope of $0.040\pm 0.004$ s.}
\label{fig_duration}
\end{figure}

Figure \ref{fig_duration}(a) presents the distribution of on-times $T$
(event durations).  The distribution is narrow and asymmetric,
yielding a mean on-time $\avT=0.124\pm 0.001$ s and standard deviation
$\sigma_T=0.051\pm 0.001$ s.  The corresponding Gaussian distribution
is depicted by the dotted line, highlighting the anomalous shape of
the measured distribution. The more moderate decrease to the right of
the peak fits an exponential decay rather than a Gaussian one
(inset). The narrow $T$ distribution is surprising in view of the
broad $l$ distribution; one expects larger events to last longer. This
correspondence is verified in Fig.\ \ref{fig_duration}(b), where the
average on-time $T_l$ is plotted as a function of the translation
$l$. Yet, the increase of $T_l$ with $l$ is only logarithmic and,
therefore, even very large translations do not have correspondingly
long duration.

The broad distribution of translations can arise from two alternative
scenarios: each observed event could correspond to either a single
fold, whereby the folds have a broad distribution of sizes, or a cascade
of roughly identical folds, the cascades having a broad
distribution of magnitudes. The following analysis, as well
as the handful of folds captured in the field of view, 
strongly support the latter scenario.

Let us assume that an observed event is caused by a
cascade of $n$ folds, each contributing roughly the same translation
$l_1$ and having an angle $\theta_i$, $i=1\ldots n$.
We assume for simplicity that the
angles are drawn from independent Gaussian distributions having a
standard deviation $\sigma_{\theta 1}$,
$
  p_{\theta_i}(\theta_i) = (2\pi\sigma_{\theta 1}^2)^{-1/2}
  e^{-\theta_i^2/(2\sigma_{\theta 1}^2)}
$.
The measured translation is the sum of contributions from all
folds, 
$
  l\simeq n l_1
$,
but the observed translation angle is the
{\em average} of the individual fold angles,
$
  \theta\simeq n^{-1}\sum_{i=1}^n\theta_i
$.
The distribution of $\theta$ is therefore narrower
the larger the value of $n$,
$
  p_{n\theta}(n,\theta) = [n/(2\pi\sigma_{\theta 1}^2)]^{1/2}
  e^{-n\theta^2/(2\sigma_{\theta 1}^2)}
$.
This implies that,
regardless of the distribution of $l$, the average absolute angle
of events having the same translation $l$ should decrease as $l^{-1/2}$,
\begin{equation}
  |\theta_l| = [2\sigma_{\theta 1}^2 l_1/(\pi l)]^{1/2},
\label{eq_theta_l}
\end{equation}
which is consistent with the findings of Fig.\ \ref{fig_angle}(b).
From the fit we get $\sigma_{\theta 1}^2 l_1 = 500\pm
200$ $(^\circ)^2\mu$m.
Thus, large cascades give rise to ``focusing'' onto
small-angle translations. This effect, along with the broad
$l$ distribution, explains the large statistical weight of small
angles.

Suppose that each fold takes a time $T_1$ to complete and
another time $\tau$ to ``topple'' another fold. Consider a cascade
made of $g$ generations of topples, where each fold can topple $q$
others \cite{ft_stochastic}. The total on-time is 
$ 
T = T_1 + (g-1)\tau 
$, and the total number of folds is
$n = g$ if $q=1$, or $n=(q^g-1)/(q-1)$ if $q>1$.  From these two
equations and $n\simeq l/l_1$ we obtain a relation between the on-time
$T$ and the translation $l$,
$T = (\tau/l_1)l + T_1 - \tau$ if $q=1$, and
$T=(\tau/\ln q) \ln[(q-1)l/l_1 + 1] + T_1 -\tau$ if $q>1$.
%
%\[
% T_l=\left\{\begin{array}{ll}
%  (\tau/l_1)l + T_1 - \tau \ \ & q=1\\
%  (\tau/\ln q) \ln[(q-1)l/l_1 + 1] + T_1 -\tau & q>1.
%  \end{array}\right.
%\]
Thus, for low cooperativity ($q=1$) we expect $T$ to depend linearly
on $l$, whereas for larger cooperativity the
dependence becomes logarithmic. As seen in Fig.\
\ref{fig_duration}(b), the results are consistent with the latter,
\ie the cascades are cooperative, one fold toppling
several others. 

Assuming a fixed probability $\alpha$ to topple a generation of
folds, we get an exponential distribution of $g$, 
$ 
p_g(g) = [(1-\alpha)/\alpha]\alpha^g 
$, 
and hence an exponential distribution of $T$ for $T>T_1$.  The
statistics for $T<T_1$ are determined by the scatter of $T_1$, which
has been neglected so far. Hence, $p_T(T)$ should have an asymmetric
shape, dropping sharply for $T<T_1$ and decaying
exponentially for $T>T_1$. These conclusions agree with the
measurements of Fig.\ \ref{fig_duration}(a), whereupon the
distribution peak can be identified as the single-fold time,
$T_1=0.12\pm 0.03$ s. From the first two moments of $p_T(T>T_1)$
we extract $\tau=0.026\pm 0.005$ s and $\alpha=0.54\pm 0.05$.
Hence, $\avg=1/(1-\alpha)=2.2\pm 0.2$, \ie the cascades consist
of 2--3 generations on average.
From the linear fit in Fig.\ \ref{fig_duration}(b), 
whose slope is equal to $\tau/\ln q$, we get 
$q=1.9\pm 0.2$, \ie each fold topples about two others.
The cooperativity and short toppling time account for the uniformity of
event duration---even the largest cascades involve only a few
generations and do not last long.

Using the exponential distribution of $g$ and the relation between $g$
and $n$, we find
\begin{equation}
p_n(n)\sim [(q-1)n + 1]^{-\beta},\ \ \beta=1-\ln\alpha/\ln q.
\label{eq_pn}
\end{equation}
The distribution of magnitudes tends for large $n=l/l_1$
to a power law, the decay exponent being always smaller than $-1$. (For
our system we get $\beta=2.0\pm 0.2$.) Using the expressions for
$p_n(n)$ and $p_{n\theta}(\theta)$ we can calculate the distribution
of angles as $p_\theta=\sum_{n=1}^\infty
p_n(n)p_{n\theta}(\theta)$, yielding
\begin{equation}
  p_\theta(|\theta|)=
  \cN_\theta\sum_{n=1}^\infty [(q-1)n+1]^{-\beta} n^{1/2}
  e^{-n\theta^2/(2\sigma_{\theta 1}^2)},
\label{eq_p_theta}
\end{equation}
where $\cN_\theta=2(2\pi\sigma_{\theta
  1}^2)^{-1/2}(q-1)^{\beta}/\zeta[\beta,q/(q-1)]$ is a normalization
factor. This distribution reproduces well the focusing of the angular
distribution, as shown by the solid line in Fig.\ \ref{fig_angle}(a).
We note that, if one has in a certain system $1<\beta<3/2$ (\ie
$\ln\alpha/\ln q>-1/2$), Eq.\ (\ref{eq_p_theta}) predicts an
(integrable) singularity of $p_\theta$ at $\theta\rightarrow 0$.  The
fit in Fig.\ \ref{fig_angle}(a) gives $\sigma_{\theta 1}=16\pm 2^\circ$
which, together with the estimate for $\sigma_{\theta 1}^2l_1$
obtained from Fig.\ \ref{fig_angle}(b), yields $l_1=2.0\pm 0.8$
$\mu$m.  Finally, we use the values derived for $l_1$, $q$ and $\beta$
to reproduce the translation distribution $p_l$ according to Eq.\ 
(\ref{eq_pn}).  As shown by the solid line in Fig.\ 
\ref{fig_offtime_trans}(a), the calculated distribution gives a reasonable
fit to the decaying part of $p_l(l)$. (The sharply increasing part is
probably determined by the scatter of $q$.) 

The cascade analysis provides a consistent account of the measured
statistics. The alternative single-fold scenario, corresponding to
$n=g=1$, does not agree with the measurements and cannot account for
the sharp distributions of angle and duration.  The evidence for
cooperative cascades, however, remains indirect. Our analysis implies
that the folding transition follows unusual nucleation kinetics, in
which single-fold growth is macroscopic in one dimension (length) but
restricted in another (width). Consequently, a single nucleus cannot
fully relax its super-stressed environment, thereby driving the
nucleation of other folds in a chain-reaction manner. Such a process,
resembling an autocatalytic chemical reaction or nuclear fission, has
never been recognized, to the best of our knowledge, in the context of
mechanical instability such as the buckling discussed here. Moreover,
this scenario should not be restricted to our specific system but is
to be expected whenever there is an autocatalytic instability whose
evolution is limited for some reason to discrete units of relaxation.

A key question is what sets the scale of the restricted fold growth.
Folding may introduce extra strain in the monolayer, \eg as a result
of mismatches in the rapidly folded region. This strain will
increase with fold width, eventually balancing the cohesion energy and
halting the folding. Another open issue is the mechanism of
correlation between folds in a cascade.  Toppling may be caused by the
long-range stress field emanating from the tips of a propagating fold.
This extra stress appears immediately after fold nucleation and can
account for the short toppling time inferred above.  A complicated
question to be addressed in a future publication \cite{ournext}
relates to the role of compression rate. It will be interesting to
check what happens to the folding cascades when the compression is not
unidirectional, \eg in a circular trough \cite{Dennin}.

\begin{acknowledgments}
  This work was partially supported by the University of Chicago MRSEC
  program of the NSF (DMR-0213745) and the US--Israel Binational
  Science Foundation (2002-271). The experimental apparatus was made
  possible by an NSF CRIF/Junior Faculty Grant (CHE-9816513).
  K.Y.C.L.\ is grateful for the support from the March of Dimes
  (\#6-FY03-58) and the Packard Foundation (99-1465). H.D.\ 
  acknowledges support from the Israeli Council of Higher Education
  (Alon Fellowship).
\end{acknowledgments}

%------------------------------------------------

%------------------------------------------------


\begin{thebibliography}{99}
\bibitem{ft_stochastic}
Strictly speaking, $T_1$, $\tau$ and $q$ are stochastic variables. We
assume that their distributions are much narrower than that of the
cascade sizes.
\end{thebibliography}
\end{document}